# A Framework for CSI-Based Indoor Localization with 1D Convolutional Neural Networks

Liping Wang, Sudeep Pasricha
Department of Electrical and Computer Engineering
Colorado State University, Fort Collins, CO, United States
{liping.wang, sudeep}@colostate.edu

*Abstract*—Modern indoor localization techniques are essential to overcome the weak GPS coverage in indoor environments. Recently, considerable progress has been made in Channel State Information (CSI) based indoor localization with signal fingerprints. However, CSI signal patterns can be complicated in the large and highly dynamic indoor spaces with complex interiors, thus a solution for solving this issue is urgently needed to expand the applications of CSI to a broader indoor space. In this paper, we propose an end-to-end solution including data collection, pattern clustering, denoising, calibration and a lightweight one-dimensional convolutional neural network (1D CNN) model with CSI fingerprinting to tackle this problem. We have also created and plan to open source a CSI dataset with a large amount of data collected across complex indoor environments at Colorado State University. Experiments indicate that our approach achieves up to 68.5% improved performance (mean distance error) with minimal number of parameters, compared to the best-known deep machine learning and CSI-based indoor localization works.

## I. Introduction

The rapidly growing demands for intellectual and human centric indoor services have made indoor localization an important component of today's edge, mobile and Internet of Things (IoT) devices. Indoor navigation services in large buildings such as museums, libraries, and shopping malls have been boosting user experience by delivering responsive positioning functionalities. Stable and accurate indoor localization capabilities are particularly crucial in highly sensitive indoor positioning use cases, such as human activity recognition in hospitals, and robot tracking and position calibration in modern factories. In extreme cases, for instance, hazardous indoor spaces in the interiors of a nuclear power plant, especially when accidental leaks occur, precise indoor positioning is vital for ground robots to fulfill radiation detection and mitigation tasks [1]. These and other use cases have led to growing interest in accurate indoor localization technologies.

Popular indoor localization solutions broadly fall into two categories: geometric mapping and feature pattern mapping (also known as fingerprinting) [2]. The former first measures predefined parameters like power, distance, direction observations, etc., with respect to some reference points, followed by calculating locations using geometric conversion algorithms such as triangulation. In contrast, with fingerprinting, the aim is to find matched feature patterns and detailed conversion algorithms are often unnecessary. Specifically, after the feature space is defined, through the comparison between the feature pattern collected from an unknown location and a reference pattern space, the coordinates of an unknown location can be approximated. For geometric mapping, the measurements of directional and distance information required in geometric mapping-based techniques heavily count on the Line-Of-Sight (LOS) conditions which are usually hard to satisfy in complex indoor environments. In addition, non-negligible approximations existing in the conversion algorithms lead to inevitable accuracy drop. In contrast, fingerprinting is regarded as a better way to handle such challenges for complex indoor scenarios, as the pattern matching does not necessarily need to account for LOS conditions, or require conversion algorithms.

Traditional RSSI (Received Signal Strength Indicator) based fingerprinting methods can suffer from constant fluctuations caused by multipath and shadowing effects which can sway RSSI values by up to 5 dB [3]. Thus, RSSI based fingerprinting techniques may lead to lower accuracy during indoor localization. Channel State Information (CSI) is considered an enhanced descriptor of wireless propagation and can improve the performance of current Wi-Fi location sensing technologies [4]. CSI data extracted from the physical layer (PHY) of fifth (and higher) generation Wi-Fi frames represents the frequency response of a Multiple Input and Multiple Output (MIMO) channel, and is capable of providing the sensitive parameters (e.g., magnitude attenuation and phase shifting) for capturing signal sources. The compensations of magnitudes and phases during signal transmissions corresponding to each subcarrier are available in captured CSI data, where uniquely different propagating paths can be spotted. In contrast to RSSI, even minor position changes of receiving antennas are able to be sensed through CSI analysis, thus there is a higher chance to achieve ultra-low localization errors.

While many promising recent efforts have demonstrated indoor localization with CSI data, several limitations remain. *First*, most state-of-the-art research with CSI data is restricted to localization within small and isolated rooms, where complex building interiors and dynamic environments, including human activities, seldom get considered. From our observations, CSI data received by a single antenna in a cuboidal shaped room has multipath and shadowing effects that are extremely limited and predictably concentrated. In more complex indoor environments, CSI patterns become more complicated and are composed of more diverse pattern groups due to the greater number of paths traversed by signals. We also found that the interior structures of a building contribute to significant modifications of CSI features during signal transmissions. Despite these structures determining the number of pattern groups, how to denoise each group remains an open question. We propose a novel approach for handling this phenomenon, which is described in Section IV. *Second*, prior works require knowledge of the Wi-Fi access points (APs), for example, the number and the identity of antennas. The access to understand each AP could be limited due to security regulations in practice, and it is not efficient or even possible to find every AP's specifications at the offline fingerprint data collection stage. *Third*, the accessibility of the dataset in most previous works is often not publicly available, thereby preventing other researchers from reproducing the results and improving the preceding research. Thus, a solution to enable an efficient, reproducible, and large-scale CSI-based indoor localization deployment without detailed AP knowledge is needed.

In the paper, we propose a novel, end-to-end deep learning based framework for CSI-based indoor localization. The key

contributions of this work can be summarized as follows:

- We propose a methodology that requires minimal knowledge of APs to preprocess CSI data in complex large open spaces, thereby removing the security concerns for accessing Wi-Fi facilities.
- We propose a lightweight, one-dimensional Convolutional Neural Network (1D CNN) with two channels that takes CSI magnitude and phase data to classify locations.
- We captured and utilized CSI data that has 180 reference points and 167 test points containing 617000 and 16700 samples (CSI packets), respectively, from 6 APs over a total path length of 183.296 meters. We plan to open source this dataset [5] to benefit the indoor localization community.
- Experimental evaluations of our framework with three state-of-the-art CSI and deep learning based localization frameworks demonstrate the promise of our approach.

## II. RELATED WORK

One of the earliest CSI fingerprinting based indoor localization studies with WLANs was conducted as part of the FIFS framework [6]. FIFS utilized the Bayes' theorem to find the maximum posteriori probability of a certain reference point (RP) given the knowledge of CSI fingerprints received from three APs at the online positioning stage and consequently, the estimated location is given by the coordinates of that RP. A key part of this work is applying spatial correlation of the CSI to determine the prior probabilities of RPs [6] for estimating unknown locations. However, the CSI from two locations separated by a small distance can be weakly unrelated in a large and complex indoor space, which makes FIFS unstable in this case. In addition, the distribution of test points (TPs) is not specified in this work which makes it difficult to replicate the performance of FIFS.

A few other efforts have explored CSI-based localization with deep learning models. DeepFi [7] uses a deep stacked autoencoder (SAE) to extract magnitude features of successive CSI packets from three transmitting antennas. In the offline stage, the SAE is individually pre-trained for each RP. CSI fingerprints from an unknown location are input to the pre-trained SAE, and the output is compared with the reconstructed CSI at RPs from the offline stage, thereby generating accumulated reconstruction errors for predicting the location. A notable drawback is the poor scalability caused by the number of SAEs needed, which grows with the number of RPs. A CNN classification model called CiFi is proposed in [8] where fingerprints are formed using angle of arrival (AoA) information to estimate locations, using phase difference information from two adjacent antennas based on 5GHz Wi-Fi. The authors mention how the estimation stability offered by AoA is better than magnitude values when LOS conditions are bad, which results in strongly attenuated magnitude data. However, CSI is essentially computed by modifications of the known preamble content via wireless paths [9]. The compensations of the weakened magnitude values discard useful information for determining multipath propagation that is valuable for tracing back the location of the signal source. DelFin [10] also uses 5GHz Wi-Fi data and adapts a CNN model that uses the CSI magnitudes collected in a 5-room apartment as inputs. DelFin requires only one anchor transmitter for residential and small working spaces. Although this solution is lightweight and suitable for IoT devices, the indoor environment analyzed is quite small and simple, without considering dynamic interference. OpenCSI [11] is an open source project that introduces a solution for automating CSI collection in a 3.5m × 5m. A radio map is built using a software-defined radio (SDR) on a wheeled robot as the collector to extract CSI from Long-Term Evolution (LTE) eNodeB. A CNN model is utilized with fusion of magnitude and phase information. The dataset is publicly made available, but the small space considered prevents considerations of dynamic and complex interference effects. Moreover, SDRs usually need dedicated infrastructure and the cost is also arguably high, for example, the USRP B200mini used in the project is priced at more than 1,000 USD. In SDR-Fi [12], a feed-forward neural network (FFNN) and 1D CNN models are built to utilize CSI magnitude for location estimation in an approximately 60 square meter space. However, CSI phase information, which has better anti-noise capabilities compared with magnitude, is not considered. In addition, details of all model layers and dataset collected are not provided, preventing comparative analysis.

Unfortunately, there are several other factors that can hinder the practical use of the aforementioned works. *First*, the sizes of experimental spaces are limited. The largest area considered is a 32.5m × 10m corridor in [6]. For larger and more complex indoor spaces, CSI patterns become more diverse and complex, and such patterns are not considered in these works. *Second*, experiments in these prior works are conducted in relatively static indoor environments. Dynamic factors such as human activities and complex electromagnetic interference are not well considered which is problematic in terms of the robustness and feasibility of real world deployment. *Third*, knowledge of Wi-Fi APs in practice might not be obtainable whenever facility security concerns matter. *Lastly*, it is also worth considering deep machine learning approaches that use less complicated models to reduce inference time for real-time localization on mobile and IoT devices.

The next section (Section III) describes our CSI data collection effort in large indoor environments. Section IV describes our data denoising and calibration approach. Section V provides an overview of our deep learning model for indoor localization. Section VI presents experimental results. Finally, Section VII concludes this work and also discusses some related open problems.

## III. DATA COLLECTION

### A. Building and Path Information

Wi-Fi devices have been pervasively adopted as the dominant infrastructure for indoor positioning, due to their low cost and the excellent universal accessibility. According to CISCO's forecast, the IP traffic generated by Wi-Fi and mobile devices is reaching 71 percent in 2022 globally, while about 549 million public Wi-Fi hotspots are going to be available by the end of the year [13]. The 802.11ac is reported to be the most prevalent standard of 802.11 family that will share 66.8 percent of WLAN endpoints by 2023 [14]. Hence, the 802.11ac standard is chosen to build the CSI dataset in this paper. We collected CSI data from the 802.11ac APs on the second floor of the Colorado State University (CSU) Behavioral Sciences Building (BSB). The gross area for data collection is 3608 square meters including 5 paths covered by 6 APs. The floorplan for data collection is shown in Fig. 1.

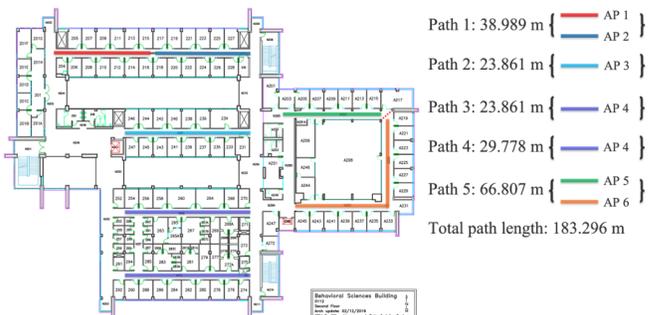

Fig. 1. The floorplan of 2nd floor of Colorado State University (CSU) Behavioral Sciences Building (BSB). Different colors represent corresponding AP visibility; each path may contain 1 or 2 APs based on received signal strength.

The data was collected during CSU operational days from Monday to Friday in a week. Collection time windows were fixed from 9am to 12 am and 2pm to 10 pm, Mountain Time (MT). The whole data collection was done over a span of two months. During the collection process, the dynamic effects and noise arising from human activities (due to students, staff, visitors, and faculty in the building) were captured and included. CSU BSB is a large modern building providing classrooms, recreation areas, labs, study rooms, conference rooms and offices where possible interference is also present during the collection windows due to the rich variety of electronic devices such as printers, plotters, computers, floor scrubbers, projectors, personal electronic devices, environmental control systems, various wireless equipment, etc.

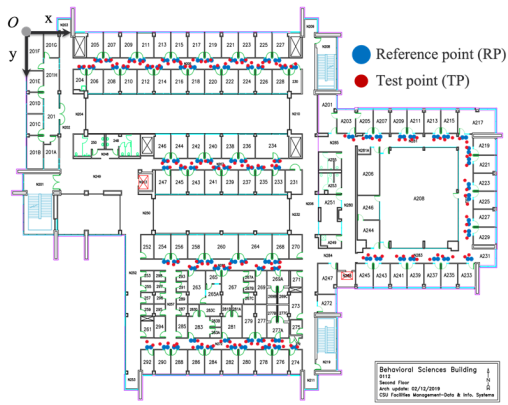

Fig. 2. The distribution of RPs and TPs on the 2nd floor of CSU BSB. RPs and TPs are annotated with blue and red, respectively.

### B. Distributions of Reference and Test Points

Fig. 2 shows the distributions of RPs and TPs where the CSI packets were collected for the building reference set and test set. To make this work practically meaningful, RPs and TPs are selected based on RSSI quality using Wi-Fi MAC scanning since the modern mobile clients can automatically select the APs with highest RSSI quality. The reference set can be further divided into training and validation set for machine learning algorithms. Every room door has two RPs that are aligned with the center of the door and the room tag of that room, respectively. TPs are selected based on our observation that if a TP is far (i.e., more than one meter) from its nearest RP, the relationship between the corresponding fingerprint of the RP and the TP tends to be weakly related, in the context of a complex indoor environment like that in the CSU BSB. This at least indicates two aspects of CSI. *First*, CSI offers impressively more sensitive information than RSSI with small location changes. *Second*, enough number of RPs with fine-grained distribution can potentially improve the accuracy of CSI-based indoor localization systems.

TABLE I: CSI collection platform components

| Hardware | Software |
|---|---|
| o Laptop | o Linux 16.04 LTS |
| o Power bank | o Raspberry Pi OS v4.19 |
| o Raspberry Pi 4 Model B | o Nexmon CSI extractor |
| o Ethernet cable | o Nexmon RSSI patch |
| o USB Type C to USB-A 2.0 charger cable | o Wi-Fi firmware version 7_45_189 for Broadcom 43455c0 chipset |
| o Laptop tripod | |

### C. Data Collection Platform

We used Nexmon [9] to extract CSI data from the PHY layer of 802.11ac symbols. Nexmon is a C-based firmware patching framework with currently the broadest support for various Broadcom and Cypress Wi-Fi chipsets. It was developed by the Secure Mobile Networking Lab (SEEMOO) with its recent version supporting 20/40 and 20/40/80 MHz per frame CSI extraction on 802.11n/ac, respectively. The data collection platform is easy and affordable to build. The hardware and software settings and peripherals are listed in Table I. The communication from the laptop to the Raspberry Pi is via an Ethernet cable where the commands are sent from the laptop depending on how the user would collect CSI packets. There is no WLAN connection between the laptop and Wi-Fi APs during the data collection. The Raspberry Pi is the only client to receive CSI packets on this platform. The height from the Raspberry Pi to the ground is fixed at 120 cm. This mean height is at an average adult's chest level, chosen based on the report from Centers for Disease Control and Prevention's (CDC) national health statistics report [15]. Here, we assume most users habitually looking at their smartphones at the height of their chest when using indoor localization services.

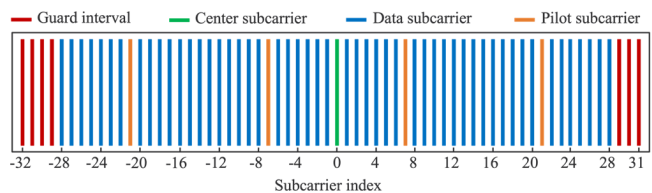

Fig. 3. A 64-subcarrier 802.11ac 20 MHz symbol

### D. 802.11ac 20MHz Symbol Structure

The 802.11ac 20MHz symbol extracted from the Nexmon CSI extractor is shown in Fig. 3. The 64 subcarriers within a single symbol includes 7 guard intervals, 52 data subcarriers, 4 pilot carriers and 1 center/null/direct current (DC) subcarrier. Guard intervals are intended to offer each segment of the preamble a cyclic delay to avoid interference [16]. Pilot carriers provide the wireless channel measurements with constellation points in data transmissions. The channel compensations from captured CSI packets are obtained based on the contents of the preamble and pilot subcarriers. The data subcarrier is the medium to carry user data and the center subcarrier is provided in 802.11ac to resist DC offset during analog/digital conversion and suppress carrier feedthrough. The values of guard intervals are constant and not useful for capturing information relevant for indoor localization, while the values of pilot and data subcarriers change based on the signal propagation paths, hence are the components we need from CSI data. The value of the center subcarrier is not used in our work and set to zero. Consequently, the magnitude and phase data from 57 subcarriers are extracted for characterizing fingerprints in this work, including 1 zero-valued center subcarrier, 4 pilot subcarriers and 52 data subcarriers.

### E. Data Collection Strategy

One major concern with indoor localization using fingerprinting is if the reference space covers enough fingerprint patterns in the area of interest and how to find these patterns. It is not easy to find the CSI patterns in a complex free space (such as in the CSU BSB environment) if there is no AP knowledge available. For example, the magnitude of 3500 successively received CSI packets at location "204_1" (with 21 abnormal packets that contain extreme values removed) is shown in Fig. 4. The magnitude computation and abnormal packet removal are explained in section IV. At first glance, from Fig. 4, there are no obvious clues showing pattern information in the randomly received packets. The challenge here is to find appropriate received signal patterns and determine the number of packets needed for the patterns at a certain location, to form fingerprints. In previous works (discussed in Section II), if the hardware and/or software stack of an AP is accessible and modifiable, at least two things are feasible to do: (1) one can inject frames over that AP to the receiver at a certain time window to avoid signal interference; (2) by streaming packets on different antennas one by one, with different antenna angles, the CSI patterns from each antenna are easy to find. These two aspects make receiving CSI patterns much more predictable compared to our situation where the lack of AP accessibility (e.g., due to security reasons) can be a bottleneck in the practical case of the APs located in the CSU BSB.

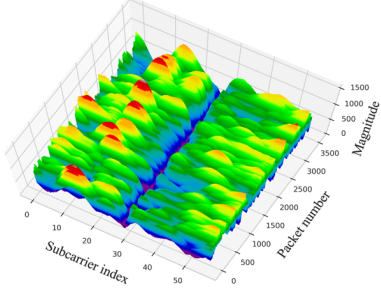

Fig. 4. Magnitude distribution of 3500 received CSI packets at the first RP of room 204 with 21 abnormal packets removed (3479 packets are visualized).

The key question now becomes: *is there a way to identify the unique signal patterns received at a specific location without any physical access to AP deployment details?* Since the arrival time of each CSI packet is highly random in our case, we hypothesize that all CSI patterns from different wireless signal paths can be determined and reproducible when the reception of packets is sufficient during an uninterrupted time window (typically over ~6 to 7 minutes per RP). To verify this hypothesis, we conducted a number of experiments and then proposed a methodology based on Monte-Carlo observations to explore the CSI patterns for complex indoor environments. In the *first* step, we gradually increased the number of received packets at each RP location and by performing a K-means clustering [17], the number of the most dominant CSI patterns were determined. In the *second* step, when the number of CSI patterns becomes stable and does not increase with the number of received packets within a 200 packet receiving window, the total number of packets for characterizing this location is determined. In the *third* step, when the number of CSI packets is determined (to include all patterns at a location from step 2), we directly set that number as the starting number for receiving the packets for the next location. This significantly reduces the complexity of the process for finding the number of patterns for each RP. We found that 2800 packets per RP was a reasonable number to capture all patterns for all RP locations in Fig. 2. To avoid random errors, extra packets are captured for each location. Thus, we finally set 3500 as the number of packets for each RP, except for the RPs covered by AP2 which have 3000 packets from each RP, due to the frequently interrupted connection after receiving 3000 packets. Note that in the online phase, our framework only requires 1 packet (sample) at a single TP to predict the location but 100 packets per TP are collected for alleviating random errors. The next section describes the CSI data pattern denoising and clustering process that is utilized as part of our framework.

## IV. CSI DATA CLUSTERING, DENOISING, AND CALIBRATION

In this section, we propose a novel data preprocessing procedure including clustering, pattern-wise denoising, and RSSI based CSI calibration. The 3500 packets received at the location 204_1 (Fig. 4) are used as an exemplar to describe the clustering and denoising methodologies.

### A. CSI Data Introduction

In our framework, CSI magnitude and phase data are extracted to train a neural network (details in Section V) for each AP. The raw CSI is a complex number composed by a real part value and an imaginary part value, as shown in equation (1). Magnitude and phase values are computed by equation (2) and (3), respectively.

$$CSI = Re(CSI) + iIm(CSI) \quad (1)$$

$$Mag(CSI) = \sqrt{Re^2(CSI) + Im^2(CSI)} \quad (2)$$

$$Pha(CSI) = \sphericalangle tan^{-1}\frac{Im(CSI)}{Re(CSI)} \quad (3)$$

The extracted CSI magnitude and phase data are visualized in Fig. 5. Note that the phase data cannot be directly used due to the discontinuities introduced by the $tan^{-1}(\cdot)$ function (Fig. 5 (b)). To solve this issue, we apply a phase unwrapping technique. If the difference between two consecutive subcarriers is equal or larger than $\pi$, the following operation is performed on the phase data along the subcarrier axis:

$$Pha(subcarrier_{i+1}) = Pha(subcarrier_i) \pm 2\pi \quad (4)$$

where $i$ represents the $ith$ subcarrier. If the adjacent difference is less than $\pi$, the original phase value of the $ith$ subcarrier is maintained. $+$ or $-$ depends on whether the difference is larger than $\pi$ or less than $-\pi$. The unwrapped phase data (which carries phase shift information inside each packet received from different wireless paths) is converted into radius as shown in the Fig. 5(c).

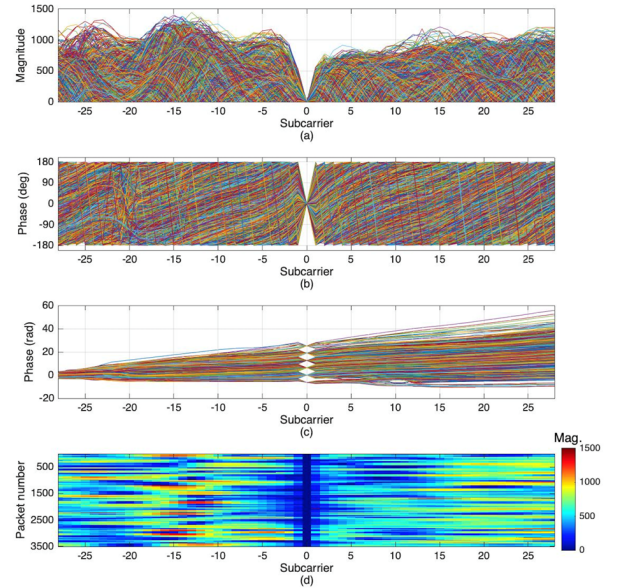

Fig. 5. Extracted CSI magnitude and phase from 3479 packets at location 204_1 with 21 abnormal packets removed. (a) Magnitude; (b) Phase in degree; (c) Unwrapped phase in radius; (d) Magnitude spectrum corresponding to Fig. 4.

### B. Pattern Clustering and Denoising

We devised a data preprocessing methodology that includes pattern clustering and denoising (discussed in this subsection) and RSSI-based CSI data calibration (discussed in the next subsection).

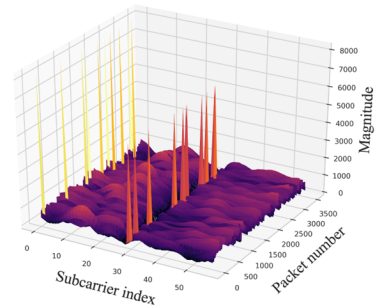

Fig. 6. Extreme magnitude values (spikes) exist in CSI packets.

*First*, the packets that have abnormal values are removed. The raw CSI data consists of extreme values that randomly appear like spikes on some subcarriers of certain packets. The magnitude of peak values is often larger than 2000 as shown in Fig. 6. Thus, 2000 is set as the threshold to remove abnormal packets for all RPs. For example, at the location 204_1, 21 out of the total 3500 received packets exceeded this threshold and were removed. *Second*, a pattern-wise denoising algorithm is utilized based on K-means clustering results. The number of clusters (i.e., number of the most dominant patterns) is determined by computing the average silhouette score (SS) based on the results of K-means. SS is obtained by the following formula:

$$SS = \frac{1}{N}\left[\sum_{i=1}^{N}(D_{inter_N} - D_{intra_i})/MAX(D_{intra_i}, D_{inter_N})\right] \quad (5)$$

where $D_{intra_i}$ is the mean distance between each signal within a cluster and $D_{inter_N}$ is the mean distance between all clusters. $i$ represents the $i$th cluster and $N$ ($N \geq 2$) is the number of clusters to evaluate based on the results of the chosen clustering method. If SS is close to 0, either the clustering algorithm does not work well or there are no distinct differences to isolate the data of interest. The number of clusters that gives the highest SS will be the number to guide our K-means algorithm.

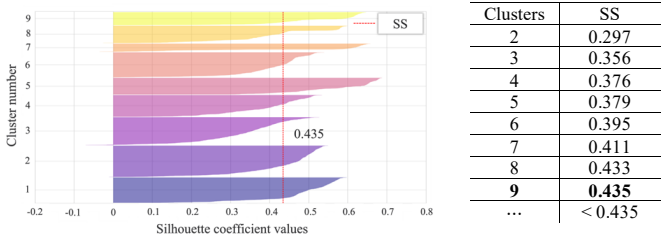

Fig. 7. The average silhouette score given by 9 clusters

As an example, Fig. 7 shows that 9 clusters give the highest SS which is 0.435 for RP location 204_1. Note that different RPs can have different clusters, based on this SS analysis. The terms cluster and dominant pattern will be used interchangeably in the following sections (e.g., cluster 1 represents dominant pattern 1 and so on).

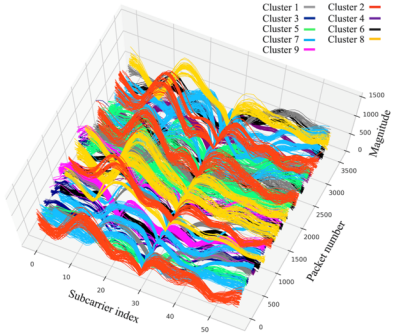

Fig. 8. The 9 magnitude patterns of 3479 CSI packets at RP location 204_1.

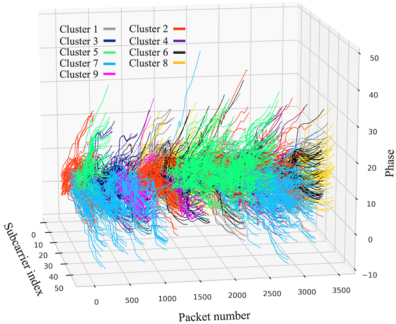

Fig. 9. The 9 phase patterns of 3479 CSI packets at RP location 204_1.

*Third*, K-means is performed with the cluster count of 9 after SS testing. The clusters from magnitude data are shown in Fig. 8. The packet indices of each cluster can be directly mapped to the indices for clustering phase data, thereby consistency between magnitude and phase can be achieved. The clustered phase patterns are shown in Fig. 9. The clustering step here is synced with the data collection process and hence there is no extra work needed since the clustering results are also used to determine the number of packets to be collected for each RP, as discussed in Section III.E. Note that we also explored creating clusters starting from phase data, but found that phase data was more stable across RPs than magnitude data, and led to the creation of fewer clusters. Thus, phase data was less effective in creating unique fingerprints for RPs than magnitude data, which is why we selected magnitude data to create clusters. However, considering both magnitude and phase patterns was crucial to achieving higher performance, which is why we use both as inputs to our deep learning model (see Section V).

*Fourth*, after finding the dominant patterns (clusters) of received CSI packets, we found the presence of dynamic noise (e.g. small-scale fading) inside each pattern. Based on our experiments, the noise tends to obfuscate the identity of RPs, thus we propose a three-stage pattern-wise denoising method, as follows:

1) Compute the mean value of each subcarrier in each pattern to determine a mean CSI sequence that represents the main feature of this pattern that the most packets contribute to.

2) Compute the correlation coefficient (CC) between the CSI packets and the sequence obtained in 1) and remove the packets that have CC values $< \psi$ (CC filtering).

3) Compute the Root Mean Squared Error (RMSE) between the remaining packets after CC filtering in 2) and the sequence obtained in 1) and remove the packets that have RMSE $> \chi$ (RMSE filtering).

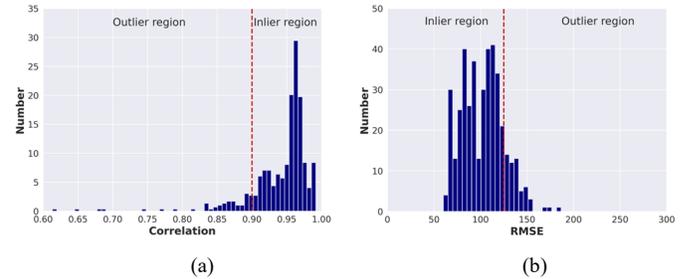

Fig. 10. (a) CC score histogram and (b) RMSE distance histogram, for pattern 2 at the location 204_1 w.r.t the mean sequence.

We empirically set the value of threshold $\psi$ to 0.9. The optimal value of threshold $\chi$ can vary from pattern to pattern inside each RP. For simplicity, we set threshold $\chi$ to 125 globally which provided good performance. The goal of the 3-stage denoising is to first filter out the most unrelated noisy packets and then the packets that are far from the mean sequence. Fig. 10 shows an example of these thresholds for cluster 2, where packets having the CC scores $< 0.9$ with respect to the mean sequence, are removed (Fig. 10(a)) and packets with RMSE $> 125$ (Fig. 10(b)) are also removed. Fig. 11 shows the result of denoising for cluster 2. The indices of removed packets are recorded to drop the corresponding phase packets directly.

## C. RSSI based CSI Calibration

Any collected CSI information is universally filtered by Automatic Gain Control (AGC) onboard modern Wi-Fi hardware whose purpose is to make the magnitudes of received signals from different propagating paths dynamically stable. For example, the most faded magnitude gets the biggest gain after AGC and vice versa. Properly dealing with AGC effects is still a challenge for CSI based indoor localization systems, due to the undocumented underlying algorithms deployed by manufacturers and the different CSI extraction approaches [18]. As a result, the magnitude values extracted from CSI packets lose the identity of the transmitter location in terms of distance which is a critical factor in realizing unique fingerprints for RPs and, consequently, localization accuracy. Since the distance information is obfuscated by AGC, one can integrate RSSI-derived distance information with CSI data for improving the performance of indoor localization systems. This is because RSSI is obtained before AGC while CSI is obtained after AGC. However, the 802.11 standard recommends an RSSI range between 0 to 255 but does not specify how RSSI needs to be calculated, thus vendors can have their own definitions to compute RSSI. In this work, we use a dBm equation to derive an RSSI-based scale factor to calibrate magnitude from CSI packets. We assume AGC is a linear time invariant system [19] that scales all magnitude values with a same scale factor once the RSSI is invariant during a certain time window. We propose a simple calibration equation:

$$RSSI = 10 log_{10}(\frac{P}{1mW}) \qquad (6)$$

$$\Rightarrow S = \sqrt{10^{\frac{RSSI}{10}}} \qquad (7)$$

where $P$ is the received signal power in milliwatt and $S$ represents the scale factor to multiply the real and imaginary parts in equation (1). The $\sqrt{(\cdot)}$ is a hardcoded conversion for converting power to voltage or current. There is no need to rescale phase data as the AGC effect is factored out with the division operation. In other words, phase data from CSI packets is theoretically more reliable in terms of ground truth, however, we focus on fusing magnitude and phase data for feeding into a deep learning model in this work. RSSI can be extracted per packet with the recent Nexmon RSSI patch for bcm 43455c0 chipset. For simplicity, we use an identical RSSI value for all packets received at a certain location. For example, 3500 packets received at the locations 204_1 have 3500 RSSI values but only one is picked to scale the magnitude values. To alleviate the effect from outlier values in the extracted RSSI, the median of 3500 RSSI values is chosen rather than the mean of them. In the online phase, captured CSI packets from TPs are preprocessed with the same procedure as that used for RPs in the offline phase, except for pattern-wise denoising.

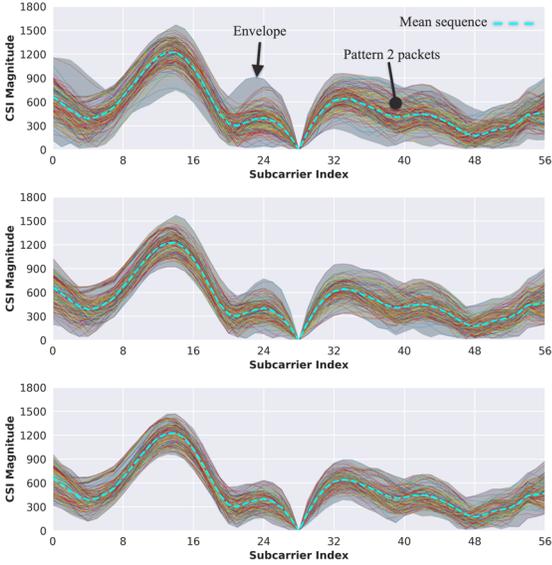

Fig. 11. The denoising process with the pattern-wise denoising algorithm. Top: the original pattern 2 packets; middle: pattern 2 packets after CC filtering; bottom: pattern 2 packets after RMSE filtering.

## V. 1D-CNN BASED LOCALIZATION

### A. Training Set and Testing Set

After the clustering, denoising, and calibration, the CSI packets are used to train a neural network. For the CSU BSB indoor environment, the number of packets (samples) used in this stage, for each AP, are listed in Table II. The samples for training from each AP are further divided into training set and validation set by the ratio of 9:1. As mentioned earlier, in the test phase, 100 samples are collected for each TP to avoid random errors. We randomly select one sample from the 100 samples for each TP, for a single test and drop it from the original set and repeat this process 10 times to compute the average distance error over 10 tests per TP. In total, 476028 and 52892 samples are used to build the training set and validation set, respectively, with the corresponding number of packets used for each AP shown in the Table II. 1670 test samples are randomly chosen from the collected 16700 CSI samples at 167 TPs with 10 samples per TP.

TABLE II: Samples for each RP after clustering/denoising/calibration

|      | AP1    | AP2    | AP3    | AP4     | AP5    | AP6    |
|------|--------|--------|--------|---------|--------|--------|
| Pkt. | 77,558 | 67,262 | 89,190 | 168,738 | 42,616 | 83,556 |

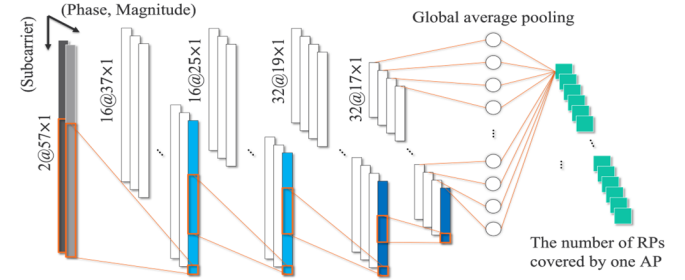

Fig. 12. The illustration of 1D CNN for location classification

TABLE III: 1D CNN layer configuration

| Layer Type | Layer Size | Filter Count | Filter Size | Stride Value | Output Size |
|---|---|---|---|---|---|
| Input | — | — | — | — | 1×57×2 |
| Convolutional | 1×37 | 16 | 21×1 | 1 | 1×37×16 |
| Dropout | Dropout rate = 0.25 | | | | |
| Convolutional | 1×25 | 16 | 13×1 | 1 | 1×25×16 |
| Dropout | Dropout rate = 0.25 | | | | |
| Convolutional | 1×19 | 32 | 7×1 | 1 | 1×19×32 |
| Dropout | Dropout rate = 0.25 | | | | |
| Convolutional | 1×17 | 32 | 3×1 | 1 | 1×17×32 |
| Dropout | Dropout rate = 0.25 | | | | |
| Global average pooling | 1×32 | 32 | 17×1 | — | 1×32 |
| Softmax | N | — | — | — | 1×N |

### B. Network Architecture

Our proposed one-dimensional (1D) CNN architecture is illustrated in the Fig. 12. The reason for selecting a 1D CNN is because 1D CNNs are not only capable of extracting features from sequence-like data, such as magnitude and phase data inside CSI packets, but can also deliver a lightweight deep neural network architecture capable of fast inferencing and low-energy consumption requirements, making it attractive for mobile devices with resource constraints. The input of this ID CNN is formed with a two-channel sequence consisting of magnitude and phase data for each channel, respectively. Details of the network architecture are shown in Table III. Each of the first two convolutional layers have 16 feature maps followed by the two convolutional layers that both have 32 feature maps. No pooling layers are involved, and all convolutional outputs are not zero-padded. We use a global average pooling layer [20] to expand the final convolutional layer. The shape of the final output layer is determined by the number of RPs covered by an AP. In the output layer, the unknown location's fingerprint is approximated to (i.e., classified as) one of the RPs within a single AP with the highest probability. Each convolutional layer is followed by a dropout layer to alleviate overfitting. The activation function is "ReLU" for each convolutional layer. Categorical cross entropy function is adopted for backpropagating classification errors during training. The number of neurons in the output layer is denoted by $N$ which is determined by the number of RPs belonging to a single AP. The value of $N$ varies from 16 (for AP5) to 56 (for AP4).

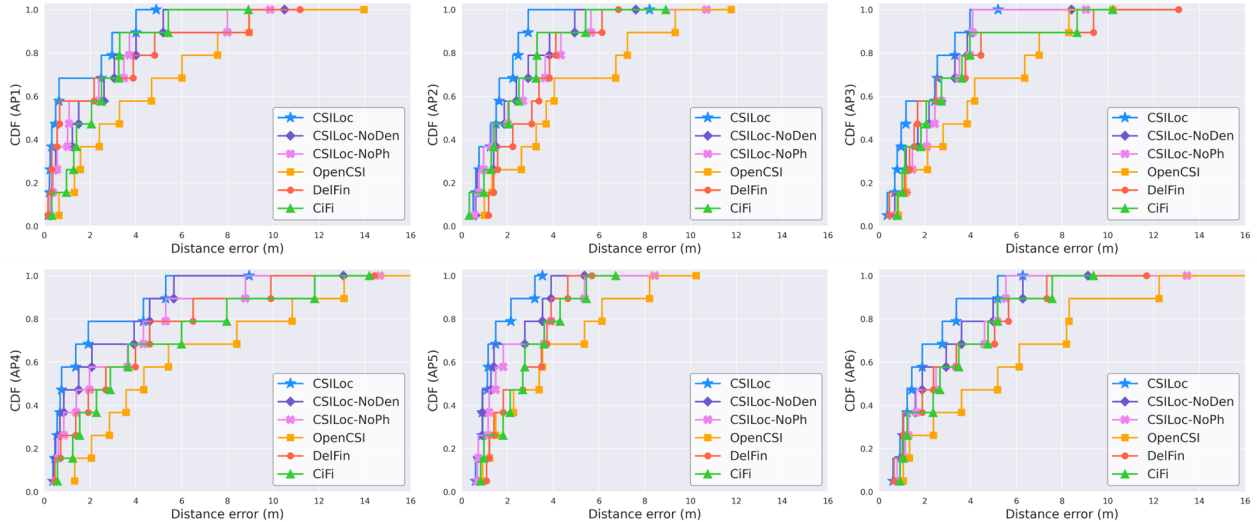

Fig. 13 Comparison of the localization errors based on CDF for each AP

## VI. EXPERIMENTAL RESULTS

We compare our model against three recent deep learning-frameworks that utilize CSI fingerprinting: CiFi [8], DelFin [10], and OpenCSI [11]. DelFin and OpenCSI are regression-based 2D CNN and 1D CNN models, respectively, with one output node for estimating the horizontal coordinate and another for the vertical coordinate of an unknown location. CiFi is a 2D CNN classification-based network with the number of output nodes being equal to the number of RPs. We compare these frameworks to three variants of our framework: a baseline variant that includes all the stages described in Sections IV and V (*CSILoc*), a variant that does not include the denoising preprocessing discussed in Section IV (*CSILoc-NoDen*), and a variant that has the same preprocessing stages as *CSILoc* but trained with only CSI magnitude data, without considering CSI phase data (*CSILoc-NoPh*), similar to the approach used in SDR-Fi [12].

Fig. 13 shows the indoor localization performance based on Cumulative Distribution Function (CDF) across the APs in the CSU BSB environment. From the results, our *CSILoc* framework shows the best distance accuracy performance across the 6 APs. The *CSILoc-NoPh* variant, which ignores phase information and only considers magnitude information performs the worst out of all the *CSILoc* variants, highlighting the importance of considering both CSI phase and magnitude information for localization. The *CSILoc-NoDen* variant outperforms the models from prior studies, although it has slightly worse performance than *CSILoc* for all APs. The superior results with *CSILoc* compared to *CSILoc-NoDen* highlights the importance of the preprocessing performed in our framework. CiFi has comparable performance with DelFin in AP1, AP3 and AP6, but is less accurate in other APs. Our analysis indicates that OpenCSI suffers from a severe model overfitting phenomenon since it is originally devised for LTE signals which have more subcarriers (higher resolution), and thus a larger number of feature maps are applied in the convolutional layers for more powerful feature extraction. As one can expect, the worst distance error results occur under AP6 for all models because the TPs within AP6 have a much larger average distance to the RPs of AP6. Compared with the results from other APs, the classification result for AP6 from our model shows a phenomenon that the neural network tends to predict more TPs as the RPs near them but not necessarily closest to them. This is understandable as the CSI fingerprinting offers more distance-sensitive features that can be potentially utilized for high accuracy indoor localization than RSSI can do. The higher average distance between RPs and TPs under AP6 makes the collected CSI signals from those RPs and TPs share less feature similarities with even short distances. The largest distance errors from OpenCSI, CiFi and DelFin come in AP4 due to more misestimated locations with the increasing number of RPs while *CSILoc* and *CSILoc-NoDen* still managed to obtain low prediction errors, highlighting their ability to scale to larger and more complex indoor environments.

Table IV: Network parameters and inference time (IT)

|            | OpenCSI    | CiFi             | DelFin | *CSILoc*        |
|------------|------------|------------------|--------|-----------------|
| Parameters | 22,910,102 | 28,052-33,852    | 53,890 | 11,280-12,600   |
| IT (ms)    | 322        | 2.25             | 2.93   | 1.89            |

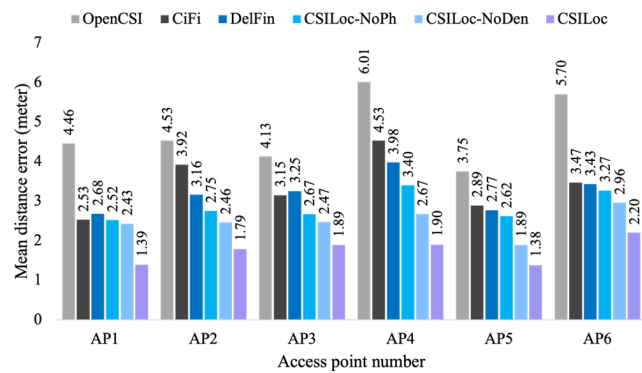

Fig. 14. Mean distance error comparison between 4 models for each AP

Fig. 14 summarizes the mean distance error for each AP, for the compared frameworks. Note that while many prior studies on CSI-based localization highlight decimeter-level accuracies, e.g., [11], these frameworks require RPs separated by centimeters which is not practical for large, real-world environments. Moreover, these studies typically consider small, isolated areas without considering dynamic interference effects over time. From the figure, it can be observed that our proposed *CSILoc* framework improves the mean distance error performance up to 68.5%, 58.1%, and 52.3% compared to OpenCSI, CiFi, and DelFin. We also obtained the inference time of all frameworks running on a Samsung Galaxy S7 smartphone with Android v8.0.0 and deep learning models prototypes with TensorFlow [21]. Table IV shows the average inference time to predict a location with *CSILoc* and the three frameworks from prior work. The table also shows the model parameters for each framework. The number of parameters for CiFi and *CSILoc* vary depending on the number of RPs, which changes the number of neurons in the final layer that are used for classification in these frameworks. To obtain inference time for these frameworks, we conservatively used the corresponding AP4 models which have the most neurons in the Softmax layer. From the table, it can be observed that our proposed *CSILoc* framework not only has the smallest memory footprint (lowest parameter count) but also the smallest inference time.

In summary, based on results shown in Fig. 13, Fig. 14, and Table IV, our proposed *CSILoc* framework shows better localization accuracy, memory footprint, and execution time, compared to state-of-the-art CSI-based localization frameworks compared against it. Thus, *CSILoc* represents a promising solution to achieve fast, lightweight, and accurate localization with CSI data in large and complex indoor environments.

## VII. Conclusion And Open Problems

In the paper, we propose a novel CSI-based indoor localization framework called *CSILoc* that involves an efficient collection, clustering, denoising, and calibration pipeline. We also proposed a 1D-CNN based neural network architecture to deliver a lightweight deep learning method towards obtaining accurate localization estimation with CSI data. Our work shows up to 68.5% improved performance (mean distance error) compared to three recent deep learning and CSI-based indoor localization frameworks. Another important contribution of our work is to create, leverage, and release an open source dataset of floor-level CSI signals collected in a large and highly dynamic indoor environment [5]. The dataset is built with a single receiving antenna client and aims at providing a lightweight dataset for resource-constrained and low-energy consumption mobile devices. This framework requires minimum knowledge about APs and does not need physical access to them (unlike frameworks in prior work that modify the APs' behavior to inject custom frames or adjust antenna configurations), which solves the problem of inaccessibility of signal sources in practice when security issues are involved in the data collection space.

Despite the promising results and the demonstrated feasibility to predict a location with CSI in a large and complex indoor environment, there are still many open problems that remain: *(i)* In our work, the number of packets in each pattern is imbalanced which introduces a bias in deep learning models, potentially limiting the generalizability of the models. New data augmentation techniques may reveal more potential from CSI data with rich multipath features; *(ii)* Our dataset is based on 802.11ac 20MHz protocol where the resolution, which depends on the number of subcarriers, is limited. The protocols with higher bandwidths such as 40MHz or 80MHz can possibly make CSI based indoor localization more sensitive and effective; *(iii)* To some extent, the performance of our framework depends on the performance of the pattern clustering algorithms. More advanced and systematic clustering algorithms can make our method more reliable; *(iv)* An optimal solution for eliminating the AGC effects is difficult to obtain, due to the undocumented AGC algorithms from chipset vendors. If such information becomes available, the effectiveness of CSI-based localization frameworks can be further improved; and *(v)* collecting enough patterns from CSI packets in a large and complex area is labor-intensive, hence, how to automate the data capture process is a critical step to make CSI-based indoor localization practically efficient and feasible. For example, when an AP of interest is updated, the CSI collected within the area covered by this AP needs to be recollected. CSI has the potential to help localization systems achieve low localization errors, however, this needs fine-grained RP allocation which means more labor is required in the data collection process.

References

[1] I. Tsitsimpelis, C. J. Taylor, B. Lennox, and M. J. Joyce, "A review of ground-based robotic systems for the characterization of nuclear environments," Progress in Nuclear Energy, vol. 111, pp. 109-124, 2019.

[2] Z. Yang, Z. Zhou, Y. Liu, "From RSSI to CSI: Indoor localization via channel response," ACM Computing Surveys (CSUR), 46:2, 2013.

[3] K. Wu, J. Xiao, Y. Yi, D. Chen, X. Luo, and L. M. Ni, "CSI-based indoor localization," IEEE TPDS, vol. 24, no. 7, pp. 1300-1309, 2012.

[4] Y. Ma, G. Zhou, S. Wang, "WiFi sensing with channel state information: A survey," ACM Computing Surveys (CSUR), vol. 52, no. 3, 2019.

[5] https://github.com/awang0413/CSILoc.

[6] J. Xiao, K. Wu, Y. Yi, and L. M. Ni, "FIFS: Fine-grained indoor fingerprinting system," in 2012 21st international conference on computer communications and networks (ICCCN), 2012: IEEE, pp. 1-7.

[7] X. Wang, L. Gao, S. Mao, S. Pandey, "CSI-based fingerprinting for indoor localization: A deep learning approach," IEEE TVT, 66:1, 763-776, 2016.

[8] X. Wang, X. Wang, and S. Mao, "CiFi: Deep convolutional neural networks for indoor localization with 5 GHz Wi-Fi," in IEEE ICC, 2017.

[9] F. Gringoli, et al., "Free your CSI: A channel state information extraction platform for modern Wi-Fi chipsets," in WNTEEC, 2019, pp. 21-28.

[10] B. Berruet, O. Baala, A. Caminada, and V. Guillet, "DelFin: A deep learning based CSI fingerprinting indoor localization in IoT context," in IEEE Intl. Conf. on Indoor Poss. and Indoor Nav. (IPIN), 2018.

[11] A. Gassner, et al., "OpenCSI: An Open-Source Dataset for Indoor Localization Using CSI-Based Fingerprinting," arXiv:2104.07963, 2021.

[12] E. Schmidt, et al., "SDR-Fi: Deep-learning-based indoor positioning via software-defined radio," IEEE Access, vol. 7, pp. 145784-145797, 2019.

[13] "Cisco visual networking index: Forecast and trends, 2017–2022," White paper, vol. 1, no. 1, 2018.

[14] "Cisco annual internet report (2018–2023) white paper," Cisco: San Jose, CA, USA, 2020.

[15] C. D. Fryar, D. Kruszan-Moran, Q. Gu, and C. L. Ogden, "Mean body weight, weight, waist circumference, and body mass index among adults: United States, 1999–2000 through 2015–2016," 2018.

[16] M. S. Gast, 802.11 ac: a survival guide: Wi-Fi at gigabit and beyond. " O'Reilly Media, Inc.", 2013.

[17] T. Kanungo, et al., "An efficient k-means clustering algorithm: Analysis and implementation," IEEE TPML, vol. 24, no. 7, pp. 881-892, 2002.

[18] M. Schulz, "Teaching your wireless card new tricks: Smartphone performance and security enhancements through wi-fi firmware modifications," 2018.

[19] I. C. S. L. M. S. Committee, "IEEE Standard for Information technology-Telecommunications and information exchange between systems-Local and metropolitan area networks-Specific requirements Part 11: Wireless LAN Medium Access Control (MAC) and Physical Layer (PHY) Specifications," IEEE Std 802.11, 2007.

[20] M. Lin, Q. Chen, and S. Yan, "Network in network," arXiv preprint arXiv:1312.4400, 2013.

[21] M. Abadi et al., "Tensorflow: Large-scale machine learning on heterogeneous distributed systems," preprint arXiv:1603.04467, 2016.